\documentclass[twocolumn,amsmath,amssymb,aps]{revtex4-2}
\usepackage{graphicx}
\usepackage{dcolumn}
\usepackage{bm}
\begin{document}

\title{ {\Large {Search for the Chiral Magnetic Effect using Sliding Dumbbell Method in Isobar Collisions ($^{96}_{44}Ru$+$^{96}_{44}Ru$ and $^{96}_{40}Zr$+$^{96}_{40}Zr$) at RHIC}}}

\author{\large Jagbir Singh for the STAR Collaboration}
\email{jagbir@rcf.rhic.bnl.gov}
\affiliation{Department of Physics, Panjab University, Chandigarh, INDIA}

\maketitle

\section*{Introduction}
In the hot dense de-confined medium in non-central heavy-ion collisions, the strong magnetic field created by the fast moving spectator protons causes the charge separation perpendicular to the reaction plane, through a phenomenon known as the Chiral Magnetic Effect (CME)~\cite{cme}.
The charge separation has been studied by using the CME sensitive $\gamma$-correlator ($\cos(\phi_a+\phi_b-2.\Psi_{RP})$)~\cite{gamma} at both RHIC and LHC energies.
It was realized that four more protons in $^{96}_{44}Ru$+$^{96}_{44}Ru$ collision than that of $^{96}_{40}Zr$+$^{96}_{40}Zr$ collision will result in larger magnetic field in Ru+Ru collision with similar backgrounds in both leading to a larger charge separation in the Ru+Ru collision~\cite{RuZr}. We compare results on charge separation in isobaric collisions at $\sqrt{s_\mathrm{NN}} = 200$ GeV.

\section*{Sliding Dumbbell Method}

\begin{figure}[!htbp]
  \centering
  \includegraphics[scale=0.20]{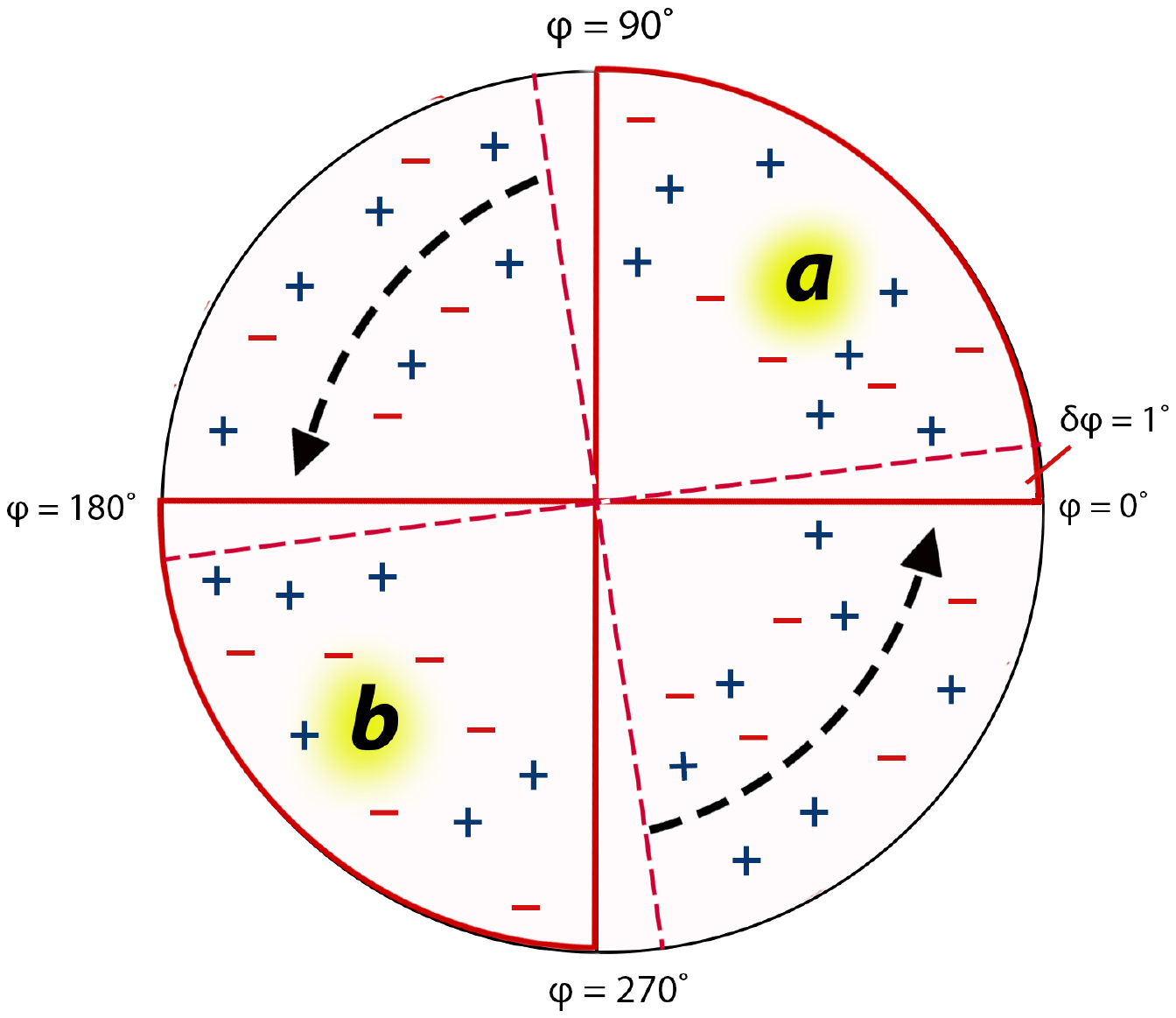} 
  \caption{Pictorial depiction of transverse (azimuthal) plane with hits of positive (+) and negative (-) charge particles in an event.}
  \label{fig:sdm}
\end{figure}

A new technique called the Sliding Dumbbell Method (SDM)~\cite{sdm} is designed to search minutely for the back-to-back charge separation on an event-by-event basis (Fig.~\ref{fig:sdm}). The azimuthal plane in each event is scanned by sliding the dumbbell of 90$^{\circ}$ in steps of $\delta\phi = 1^{\circ}$, while calculating $Db_{+-}$ for each region, to obtain maximum values of $Db_{+-}$ ($Db^{max}_{+-}$) in each event. $Db_{+-}$ is the sum of positive and negative charge fraction on ``a'' and ``b'' side of dumbbell, respectively (Eq.~\ref{eq:db}).
\begin{equation}
  \label{eq:db}
  Db_{+-} =  \frac{n^a_{+}}{(n^a_{+}+n^a_{-})} + \frac{n^b_{-}}{(n^b_{+}+n^b_{-})},
\end{equation}
where $n^a_{+}$ and $n^a_{-}$ ($n^b_{+}$ and $n^b_{-}$) are numbers of positive and negative charged particles on “a” (“b”) side of the dumbbell. The fractional charge separation ($f_{DbCS}$) across the dumbbell in each event is defined as:

\begin{equation}
  \label{eq:fdbcs1}
  f_{DbCS} = Db_{+-}^{max} -1. 
\end{equation}

For the background estimation, the charges of particles in each event are shuffled randomly to destroy the charge dependent correlations amongst the charged particles, keeping $\theta$ and $\phi$ of each particle unchanged in an event. This is called the charge shuffled event sample. However, to recover correlations amongst particles which were destroyed in charge shuffling, we obtain the correlations from the original events corresponding to the particular $f_{DbCS}$ bin of charge shuffeld events, which is termed as the correlated background.

\section*{Results and Discussions}
\begin{figure}[!htbp]
  \centering
  \includegraphics[scale=0.18]{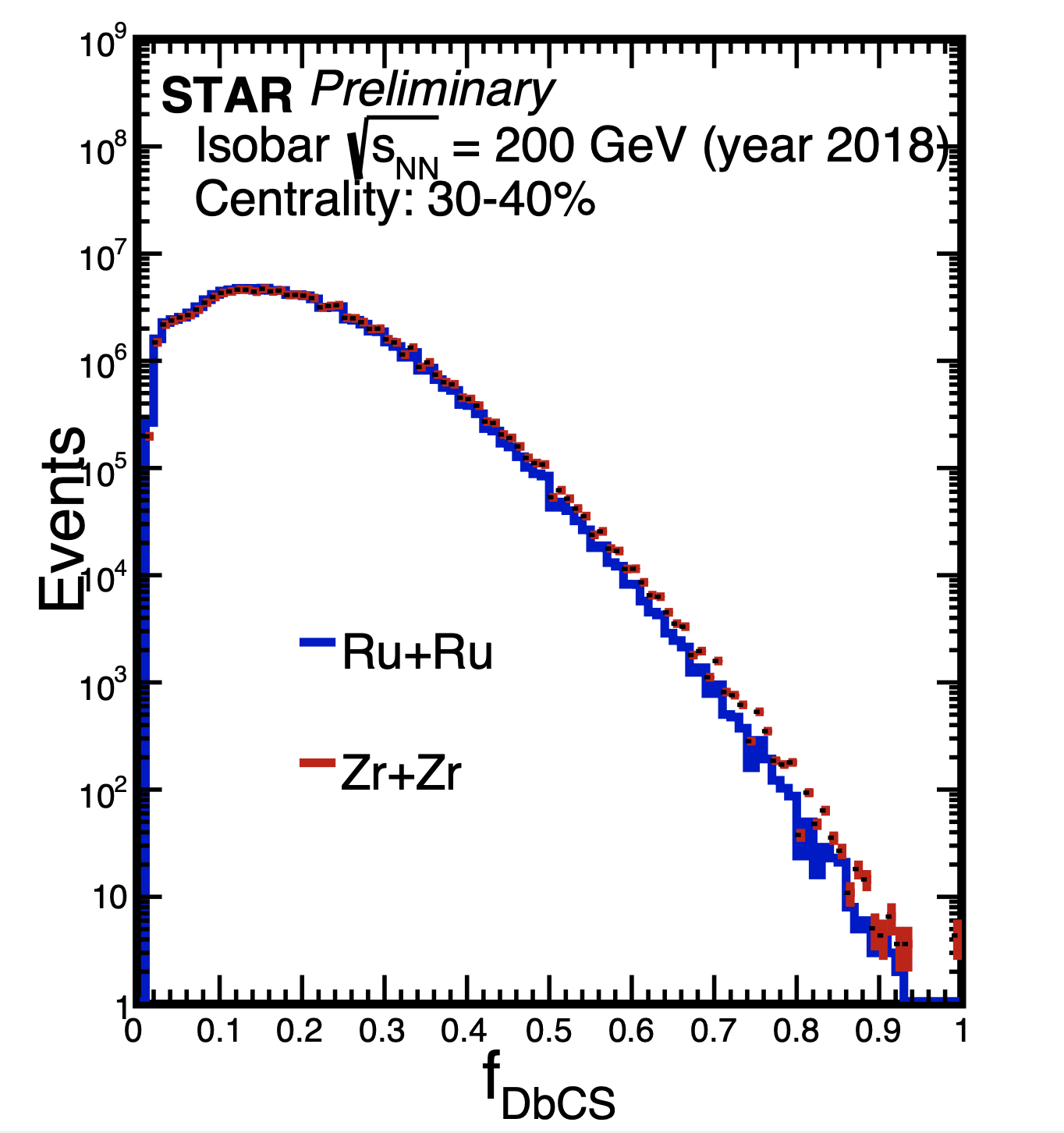}
  \caption{The fractional dumbbell charge separation ($f_{DbCS}$) distribution for 30-40$\%$ centrality for Ru+Ru and Zr+Zr collisions.}
  \label{fig:fdcs}
\end{figure}

It is seen that the $f_{DbCS}^{Zr+Zr}$ distribution is leading $f_{DbCS}^{Ru+Ru}$ distribution for 30-40$\%$ collision centrality. Different $f_{DbCS}$ distributions are  divided into ten percentile bins for each centrality.

\begin{figure}[!htbp]
  \centering
  \includegraphics[scale=0.28]{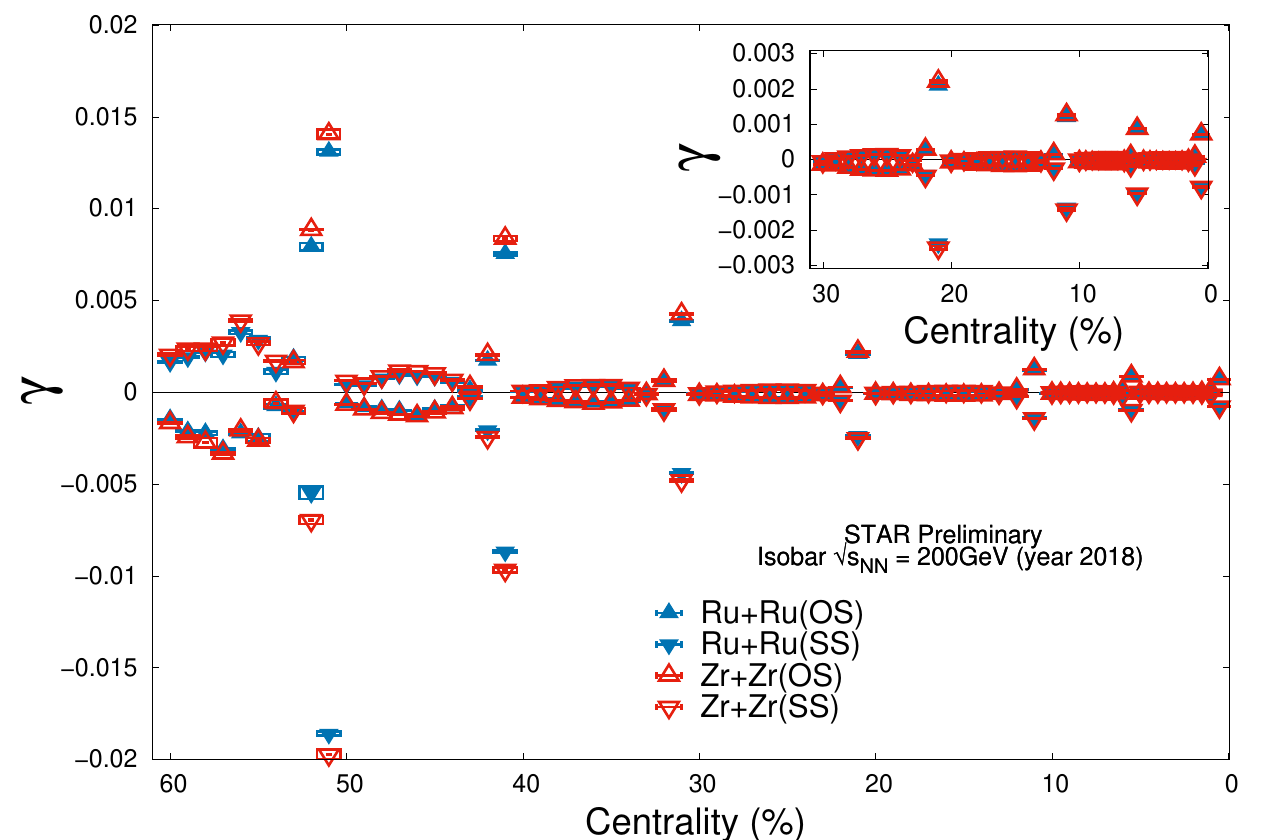}
  \caption{$\gamma_{OS}$ and $\gamma_{SS}$  as a function of $f_{DbCS}$ in each centrality for Ru+Ru and Zr+Zr collisions.}
  \label{fig:gammassos}
\end{figure}

Figure~\ref{fig:gammassos} represents $\gamma$-correlator for opposite (OS) and same sign (SS) charge pairs corresponding to ten  $f_{DbCS}$ bins in each collision centrality. It is seen that for CME-like events (those with largest $f_{DbCS}$ values), $\gamma_{OS} > 0$ and $\gamma_{SS} < 0$ for top 20$\%$(30$\%$) $f_{DbCS}$ bins for 0-40$\%$(40-60$\%$) centralities. Boxes represent the systematic uncertainties, while statistical uncertainties are smaller than the symbol size.

\begin{figure}[!htbp]
  \centering
 \includegraphics[scale=0.28]{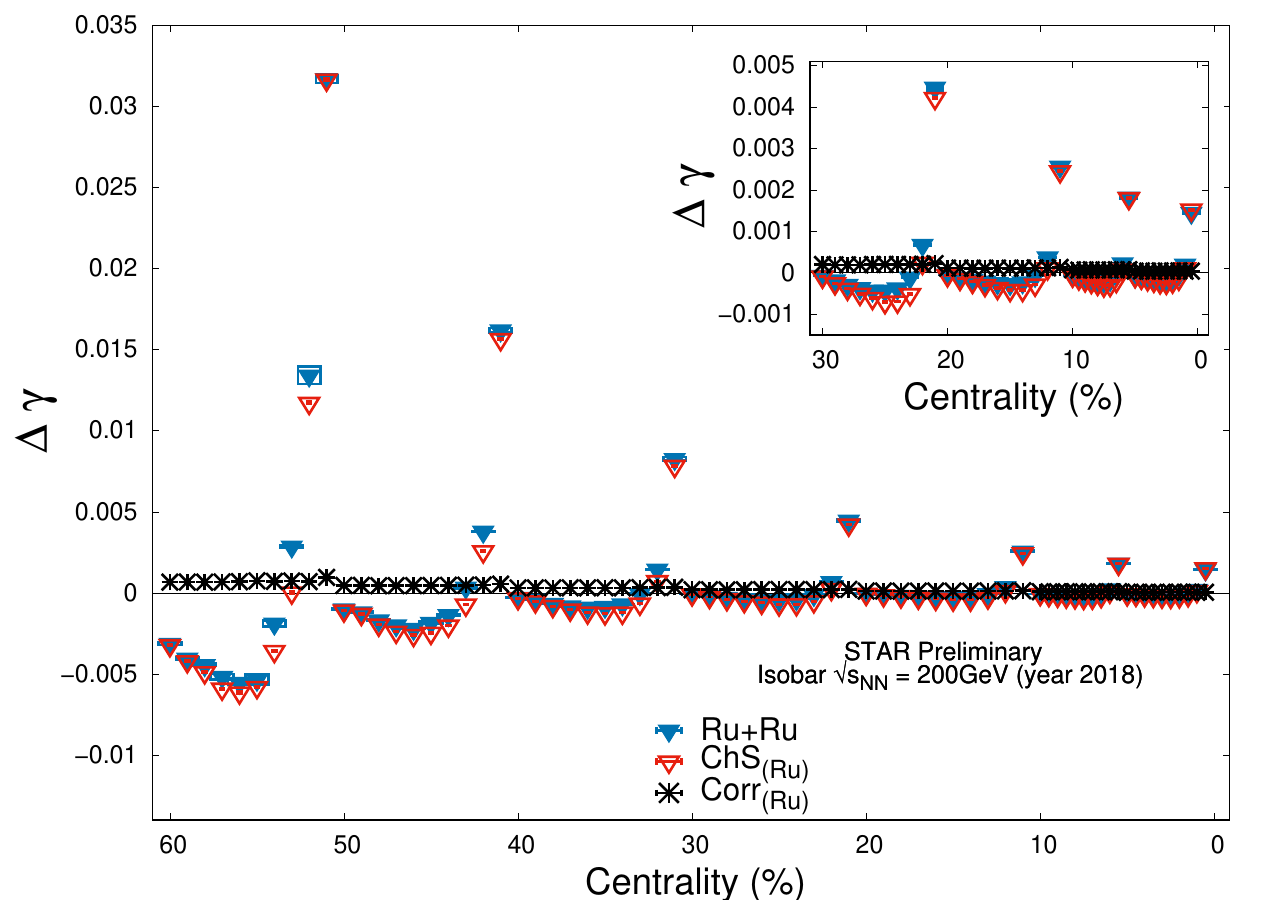}
 \caption{Comparison of $\Delta\gamma^{Ru+Ru}$ with charge shuffle and correlated background for 0-60$\%$ collision centrality.}
 \label{fig:deltagammaRu}
\end{figure}

\begin{figure}[!htbp]
  \centering
  \includegraphics[scale=0.28]{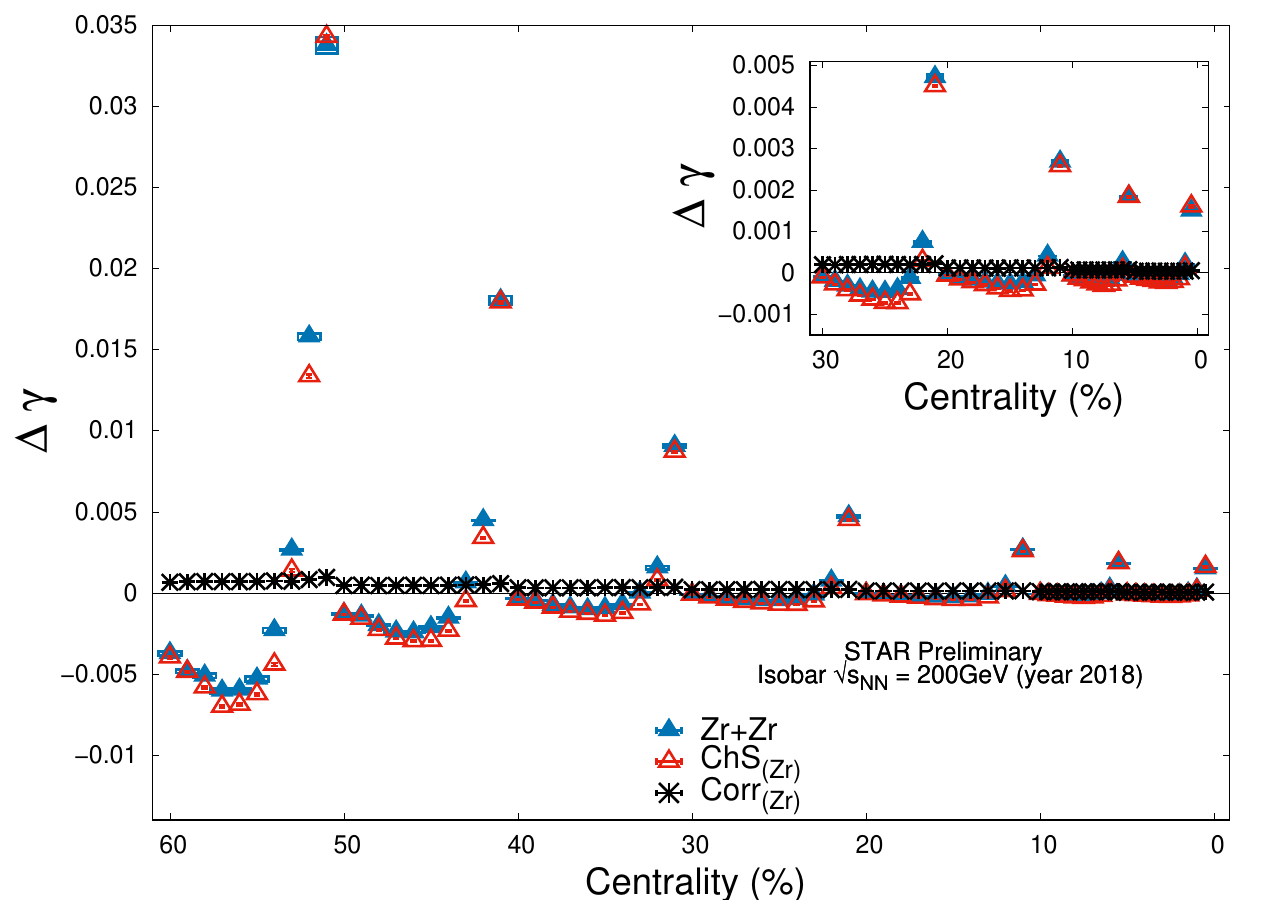}
  \caption{Comparison of $\Delta\gamma^{Zr+Zr}$ with charge shuffle and correlated background for 0-60$\%$ collision centrality.}
  \label{fig:deltagammaZr}
\end{figure}

In Figs.~\ref{fig:deltagammaRu} and~\ref{fig:deltagammaZr}, $\Delta\gamma$ for Ru+Ru and Zr+Zr are compared with Charge shuffled ($\Delta\gamma_{ChS}$) and Correlated ($\Delta\gamma_{Corr.}$) backgrounds for 0-60$\%$ collision centralities. It is seen that $\Delta\gamma$ is positive for both Ru and Zr (both for data and background i.e., ChS and Corr) in the top 20$\%$(30$\%$) $f_{DbCS}$ bins in 0-40$\%$(40-60$\%$) centralities. We are able to identify the events with higher charge separation (CME-like) corresponding to higher  $f_{DbCS}$ values in each centrality. Results based on above will be discussed in detail.



\end{document}